%% file: main_arxiv.tex
\documentclass[conference]{IEEEtran}
\IEEEoverridecommandlockouts
\usepackage{cite}
\usepackage{amsmath,amssymb,amsfonts}

\usepackage{algorithmic}
\usepackage{graphicx}
\usepackage{textcomp}
\usepackage{xcolor}

\usepackage[acronym]{glossaries}
\usepackage{scalerel}
\usepackage{tikz}
\usepackage{svg}
\usepackage{hyperref}
\usepackage{mathtools}
\usepackage{multirow}
\usepackage{siunitx}

\usepackage[a4paper, total={184mm,239mm}]{geometry}

\usetikzlibrary{svg.path}
\definecolor{orcidlogocol}{HTML}{A6CE39}
\tikzset{
  orcidlogo/.pic={
    \fill[orcidlogocol] svg{M256,128c0,70.7-57.3,128-128,128C57.3,256,0,198.7,0,128C0,57.3,57.3,0,128,0C198.7,0,256,57.3,256,128z};
    \fill[white] svg{M86.3,186.2H70.9V79.1h15.4v48.4V186.2z}
                 svg{M108.9,79.1h41.6c39.6,0,57,28.3,57,53.6c0,27.5-21.5,53.6-56.8,53.6h-41.8V79.1z M124.3,172.4h24.5c34.9,0,42.9-26.5,42.9-39.7c0-21.5-13.7-39.7-43.7-39.7h-23.7V172.4z}
                 svg{M88.7,56.8c0,5.5-4.5,10.1-10.1,10.1c-5.6,0-10.1-4.6-10.1-10.1c0-5.6,4.5-10.1,10.1-10.1C84.2,46.7,88.7,51.3,88.7,56.8z};
  }
}
\newcommand\orcidicon[1]{\href{https://orcid.org/#1}{\mbox{\scalerel*{
\begin{tikzpicture}[yscale=-1,transform shape]
\pic{orcidlogo};
\end{tikzpicture}
}{|}}}}

\newcommand{\mSIUnm}{\si{\nano\meter}}
\newcommand{\mSIUuS}{\si{\micro\siemens}}
\newcommand{\mSIUfF}{\si{\femto\farad}}
\newcommand{\mSIUfJ}{\si{\femto\joule}}

\def\preprint{1} 

\if\preprint1
\usepackage[all]{background}
\usepackage{stackengine}
\fi

\begin{document}

\title{A Calibratable Model for Fast Energy Estimation of MVM Operations on RRAM Crossbars}

\if\preprint1
\setstackEOL{\\}
\setstackgap{L}{\normalbaselineskip}
\SetBgContents{\color{gray}{\tiny \Longstack{PREPRINT - Accepted at 2024 IEEE 6th International Conference on Artificial Intelligence Circuits and Systems (AICAS)}}}
\SetBgPosition{4.5cm,1cm}
\SetBgOpacity{1.0}
\SetBgAngle{0}
\SetBgScale{1.8}
\fi

\author{
  \IEEEauthorblockN{
    José Cubero-Cascante \orcidicon{0000-0001-9575-0856},
    Arunkumar Vaidyanathan \orcidicon{0000-0003-3428-6775},
    Rebecca Pelke \orcidicon{0000-0001-5156-7072},
    Lorenzo Pfeifer \orcidicon{0009-0004-6191-6482},\\
    Rainer Leupers \orcidicon{0000-0002-6735-3033},
    Jan Moritz Joseph \orcidicon{0000-0001-8669-1225}
  }
  \IEEEauthorblockA{
    \textit{RWTH Aachen University, Institute for Communication Technologies and Embedded Systems}, Aachen, Germany \\
    \{cubero,vaidyanathan,pelke,pfeifer,leupers,joseph\}@ice.rwth-aachen.de
  }
  \thanks{This work was funded by Germany's Federal Ministry of Education and Research (BMBF) in the project NEUROTEC II (Project Nos. 16ME0398K, 16ME0399).}

  }

\newacronym[plural=MVMs,firstplural=Matrix Vector Multiplications (MVMs)]{mvm}{MVM}{Matrix Vector Multiplication}
\newacronym[plural=VMMs,firstplural=Vector Matrix Multiplications (VMMs)]{vmm}{VMM}{Vector Matrix Multiplication}
\newacronym[plural=DNNs,firstplural=Deep Neural Networks (DNNs)]{dnn}{DNN}{Deep Neural Network}
\newacronym[plural=CNNs,firstplural=Convolutional Neural Networks (CNNs)]{cnn}{CNN}{Convolutional Neural Network}
\newacronym{rram}{RRAM}{Resistive Random Access Memory}
\newacronym{pcm}{PCM}{Phase Change Memory}
\newacronym[plural=ADCs,firstplural=Analogue-to-Digital Converters (ADCs)]{adc}{ADC}{Analogue-to-Digital Converter}
\newacronym[plural=DACs,firstplural=Digital-to-Analogue Converters (DACs)]{dac}{DAC}{Digital-to-Analogue Converter}
\newacronym{soc}{SoC}{System-on-Chip}
\newacronym{cim}{CIM}{Compute-in-Memory}
\newacronym{mac}{MAC}{Multiply-Accumulate}
\newacronym{kcl}{KCL}{Kirchhoff's Circuit Laws}
\newacronym{itrs0}{ITRS}{International Technology Roadmap for Semiconductors}
\newacronym{vcm}{VCM}{Valence Change Mechanism}
\maketitle

\begin{abstract}
The surge in AI usage demands innovative power reduction strategies.
Novel Compute-in-Memory (CIM) architectures, leveraging advanced memory technologies, hold the potential for significantly lowering energy consumption
by integrating storage with parallel Matrix-Vector-Multiplications (MVMs).
This study addresses the 1T1R RRAM crossbar, a core component in numerous CIM architectures.
We introduce an abstract model and a calibration methodology for estimating operational energy.
Our tool condenses circuit-level behaviour into a few parameters, facilitating energy assessments for DNN workloads.
Validation against low-level SPICE simulations demonstrates speedups of up to 1000$\times$ and energy estimations with errors below 1\%.
\end{abstract}


\section{Introduction}

With the steep rise in the adoption of AI, its power demand surged.
Novel approaches that substantially reduce power consumption are needed to achieve a sustainable use of AI.
Architectures built around emerging \gls{cim} technologies have the potential to reduce the energy of data-intensive workloads drastically.
They mitigate the memory wall in traditional von Neumann systems, as costly data movements are reduced~\cite{cim_review_nature_2023}.

One emerging memory class is non-volatile resistive memory, such as \gls{rram} and \gls{pcm}.
It enables considerable power savings via low-energy analogue computations~\cite{review_2020_analog_arch_NN_NVM}.
Further advantages are the absence of leakage power and the multi-bit storage capability~\cite{Yu_ReviewCimChipsDL2021}.
In RRAM-based \gls{cim} systems, the memory cells are arranged in crossbars that compute \glspl{mvm},
a fundamental operation in \glspl{dnn}, in $\mathcal{O}(1)$ time complexity.
This concept was integrated into several architectures~\cite{ISAAC_2016,Puma2019,PipeLayer2017,VLSI2021_hermes_macro,FullCNNNature2020}.

Due to the considerable savings promised by \gls{cim} technology, numerous technology stacks, data representations, and peripheral circuits have been published,
leaving behind a fragmented landscape.
As established methods for performance analysis from digital CMOS cannot be directly used to evaluate the performance of analogue \gls{cim} systems,
a custom framework is missing.
A system-level evaluation of the power consumption for real-world AI algorithms is needed;
hence, the framework must yield high simulation performance.
Device and circuit-level properties, such as the geometry of transistors and wire parasitics in 1T1R arrays, strongly affect energy efficiency.
Thus, the framework must be calibrated to device-level simulations.

\begin{figure}
    \includegraphics[width=\linewidth]{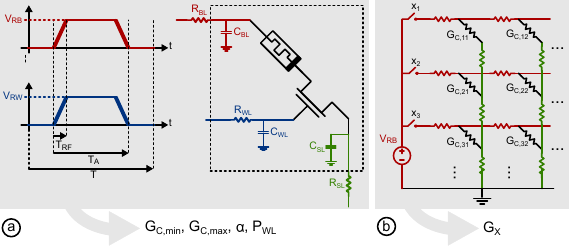}
    \caption{The Gist of Our Energy Model: (a) Cell Calibration and (b) Crossbar Circuit Reduction Including Wire Parasitics}
    \label{fig:cell_calib_gx}
\end{figure}

This work introduces the required fast and calibrated framework to quantify the combined effect of device properties,
circuit designs, and system design decisions on the energy consumed by 1T1R crossbars in CIM-based systems.
Our main contributions are:
\begin{itemize}
    \item A methodology to study the effect of the selection transistor and wire parasitics in 1T1R crossbar arrays.
    \item A comparison of the energy budget of the state-of-the-art weight mappings for crossbar-based MVM operations.
    \item A calibrated and fast system energy model for MVM operations on resistive memory crossbars, enabling the evaluation of real workloads.
\end{itemize}
Fig.~\ref{fig:cell_calib_gx} illustrates the pivotal elements driving this work: cell calibration using read pulses and circuit reduction, including wire parasitic resistances.

\section{Background}

\subsection{Resistive MVM Compute Kernels}

\begin{figure}
    \begin{centering}
        \includegraphics[width=0.95\linewidth]{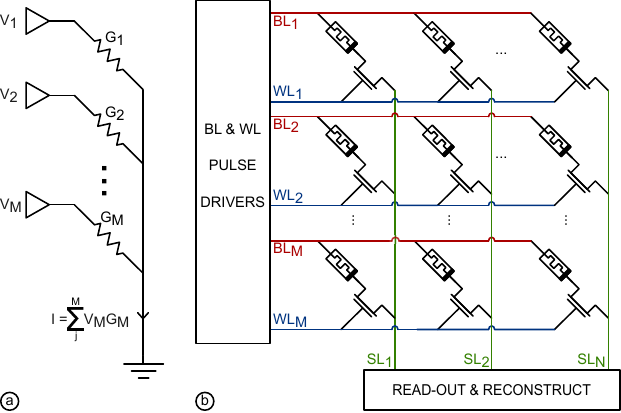}
        \centering
        \caption{Ideal Resistive Dot-product and 1T1R Crossbar}
        \label{fig:1t1rxbar}
    \end{centering}
    \end{figure}

Resistive arrays enable dot-product operations in the analogue domain~\cite{dot_product_engine_2016}, as depicted in Fig.~\ref{fig:1t1rxbar}a.
A vector $\mathbf{G} = [G_1, G_2, \ldots, G_M]$ is encoded as the conductance of each resistive device,
while the applied voltages $\mathbf{V} = [V_1, V_2, \ldots, V_M]$ represent the second vector.
According to \gls{kcl}, the output current is $I=\sum_j^M V_jG_j = \mathbf{V} \cdot \mathbf{G}$.
Each resistive element acts as storage and \gls{mac} unit simultaneously.
A passive crossbar, built by expanding columns in Fig.~\ref{fig:1t1rxbar}a, realises a fully parallel implementation of the \gls{mvm} operation.

Passive crossbars suffer from sneak-path currents, so their usage for multi-level cell storage is limited~\cite{MetalOxideRRAM_HSWong_2012}.
An established solution to this problem is providing each memristor with a transistor that acts as an access device, forming a 1T1R crossbar
(see Fig.~\ref{fig:1t1rxbar}b).
The transistor regulates the maximum current, enabling fine-tuning of individual conductances~\cite{cim_review_nature_2023}.

The integration of 1T1R cells in large crossbar arrays requires long metal wires that suffer from parasitic line resistance and capacitance.
These parasitic elements affect the accuracy of computations but also impact energy efficiency~\cite{ParasiticsCrosSim2021}.

\subsection{CNN Workloads as MVMs}

\glspl{cnn} are widely recognised and extensively utilised in various applications,
including successfully detecting, segmenting, and recognising objects and regions in images~\cite{DL_LeCun2015}.

In \glspl{cnn}, the computationally intensive tasks primarily reside in the convolution layers~\cite{Puma2019}.
To execute convolution layers on a resistive \gls{cim} kernel, they must first be translated into \glspl{mvm}~\cite{pelke2023mapping}.
The \textit{im2col}~\cite{im2col2016} transformation is a widely used technique for this purpose.

\gls{cnn} acceleration with RRAM-based \gls{cim} kernels has been demonstrated in silicon by several works~\cite{FullCNNNature2020,VLSI2021_hermes_macro,NEURRAM_Wan2022}.
The trained weights of CNN layers are mapped to memristive crossbars, while input activations are encoded to a train of voltage pulses.
Integer quantisation is usually applied to weights and activations before encoding~\cite{quantRRAM2020,quantRRAMReview2022}.

\subsection{Weight Mapping} \label{sec:weightmap}

Memristive cells can only hold positive values between $G_{min}$ and $G_{max}$.
Therefore, negative weights have to be expressed as positive values.
Furthermore, due to their intrinsic cycle-to-cycle variability,
only a finite number of states are feasible~\cite{2022_ISCAS_FundLim_CIM_EDA_RRAM}.

There are two main weight mapping approaches~\cite{Yu_ReviewCimChipsDL2021}.
With the \textit{bias mapping}, signed weights are converted to conductances using a fixed scaling factor $s$ and a bias $G_b$, so that $G_{ij}=sW_{ij}+G_b$.
The bias $G_b$ is set to $(G_{min}$+$G_{max})/2$~\cite{ISAAC_2016}.

With the \textit{differential mapping}, the weight values are represented with a pair of values, so that $sW_{ij} = G_{ij}^+ - G_{ij}^-$.
Typically, $G_{ij}^-$ is set to $G_{min}$ for positive numbers and $G_{ij}^+$ is set to $G_{min}$ for negative numbers~\cite{Sandia_Accuracy}.

After mapping, values can be programmed to one (bias) or two (differential) memristive cells.
However, if the range of the integer quantisation is greater than the device achievable states,
the value will be further split by applying \textit{cell spatial bit-slicing}~\cite{ISAAC_2016}.

\subsection{Input Encoding}

Directly mapping arbitrary multi-bit input values to input voltages is hard to realise, as the memristor IV response is often linear only in a small range\cite{bengel_bitslicing}.
Moreover, \gls{dac} and \gls{adc} circuits are costly in terms of area and power~\cite{mixedsignaldnn2021}.
A common solution is \textit{input temporal bit-slicing}: passing the input values bit-by-bit~\cite{bitslicing_said_2023}.
This method is assumed for our analysis, as it is used in a significant number of \gls{cim} chips
and architecture concepts~\cite{FullCNNNature2020,Puma2019,ISAAC_2016,NEURRAM_Wan2022,dot_product_engine_2016}.

\section{Methods}

\subsection{Cell Pulse Energy Model based on Calibration} \label{sec:cell_pulse_energy}

We propose an abstract energy model with a small number of calibrated parameters.
Our analysis starts at the 1T1R cell level, where the analogue \gls{mac} operations take place.
As shown in Fig.~\ref{fig:cell_calib_gx}a, the 1T1R cell circuit includes the memristive device, the selection transistor, and
the unit wire parasitic capacitances for the three interconnect lines: BL, WL, and SL.
The wire parasitic resistances are excluded, as their effect will be considered in the crossbar-level model in the next section.

A \gls{mac} operation is performed by applying pulses of amplitude $V_{RB}$ and $V_{RW}$ to the BL and WL lines, respectively.
The pulses feature the same total $T$, activation $T_A$, and rise-fall $T_{RF}$ times.
Both BL and WL lines are activated simultaneously if the 1-bit input $x$ is one and are kept low otherwise.
The cell energy can be expressed as:

\begin{equation} \label{eq:ec1}
    E_{C} = x(E_{WL} + E_{BL}) = xT(P_{WL} + P_{BL}).
\end{equation}

$P_{WL}$ corresponds to the average power delivered to the WL input in one pulse.
This power stems from charging and discharging the gate capacitance and the parasitic wire capacitance $C_{WL}$.
Its magnitude depends on the transistor geometry and the input pulse's parameters $T_A$ and $T_{RF}$.
However, it remains constant, disregarding the cell's resistive state.

Similarly, $P_{BL}$ is the average power flowing into the BL input in one pulse.
We define it as the product of the cell's steady-state power $P_S$ and a calibration parameter $\alpha$:
\begin{equation} \label{eq:ps}
    P_{BL} = \alpha P_{S} = \alpha V_{C}^2 G_{C}.
\end{equation}

$G_C$ is the cell's apparent conductance, which combines the memristor's state and the transistor's on-resistance $R_{TON}$.
$V_{C}$ is the voltage seen by the cell in the steady state.
The calibration parameter $\alpha$ collects the transient elements, including the transistor switching energy and the charging and discharging of
the parasitic capacitances $C_{BL}$ and $C_{SL}$.

The resulting cell pulse energy equation is:
\begin{equation} \label{eq:ec2}
    E_{C} = Tx(\alpha V_C^2G_C + P_{WL}).
\end{equation}

\subsection{Total Crossbar MVM Energy}

The crossbar's total \gls{mvm} energy can be expressed as the sum of the energy of all cells and $P$ pulses.
For a crossbar with $X_M$ columns and $X_N$ rows, this is:
\begin{equation}
    E_{MVM} = \sum_{p}^{P} \sum_{i}^{X_M} \sum_{j}^{X_N} E_{C,p,ij}.
\end{equation}
Introducing the cell pulse energy from Eq.~\ref{eq:ec2}, we get:
\begin{multline} \label{eq:emvm_long}
    E_{MVM} = \\
    T \sum_{p}^{P} (\alpha \sum_{i}^{X_M} \sum_{j}^{X_N} x_{p,j} V_{C,ij}^2 G_{C,ij} + X_M P_{WL} \sum_{j}^{X_N}x_{p,j}).
\end{multline}
With the \textit{input temporal bit-slicing} scheme, $x_{p,j}$ is a single-bit value, while $P$, $T$, and $P_{WL}$ are constants.
However, as illustrated in Fig.~\ref{fig:cell_calib_gx}b, $V_{C,ij}$ is not constant due to the parasitic wire resistances.
To address this variability, we employ the fast circuit solver from~\cite{FastParasiticsJETCAS2022} to calculate the steady-state equivalent conductance $G_{X,p}$ that satisfies:
\begin{equation} \label{eq:thevenin}
    \sum_{i}^{X_M} \sum_{j}^{X_N} x_{p,j} V_{C,ij}^2 G_{C,ij} = V_{RB}^2 G_{X,p}.
\end{equation}
Finally, the \gls{mvm} energy becomes:
\begin{equation} \label{eq:emvm_short}
    E_{MVM} = T \sum_{p}^{P} (\alpha V_{RB}^2 G_{X,p} + X_M P_{WL} \sum_{j}^{X_N}x_{p,j}).
\end{equation}

\subsection{Workflow}

\begin{figure}
\begin{centering}
    \includegraphics[width=\linewidth]{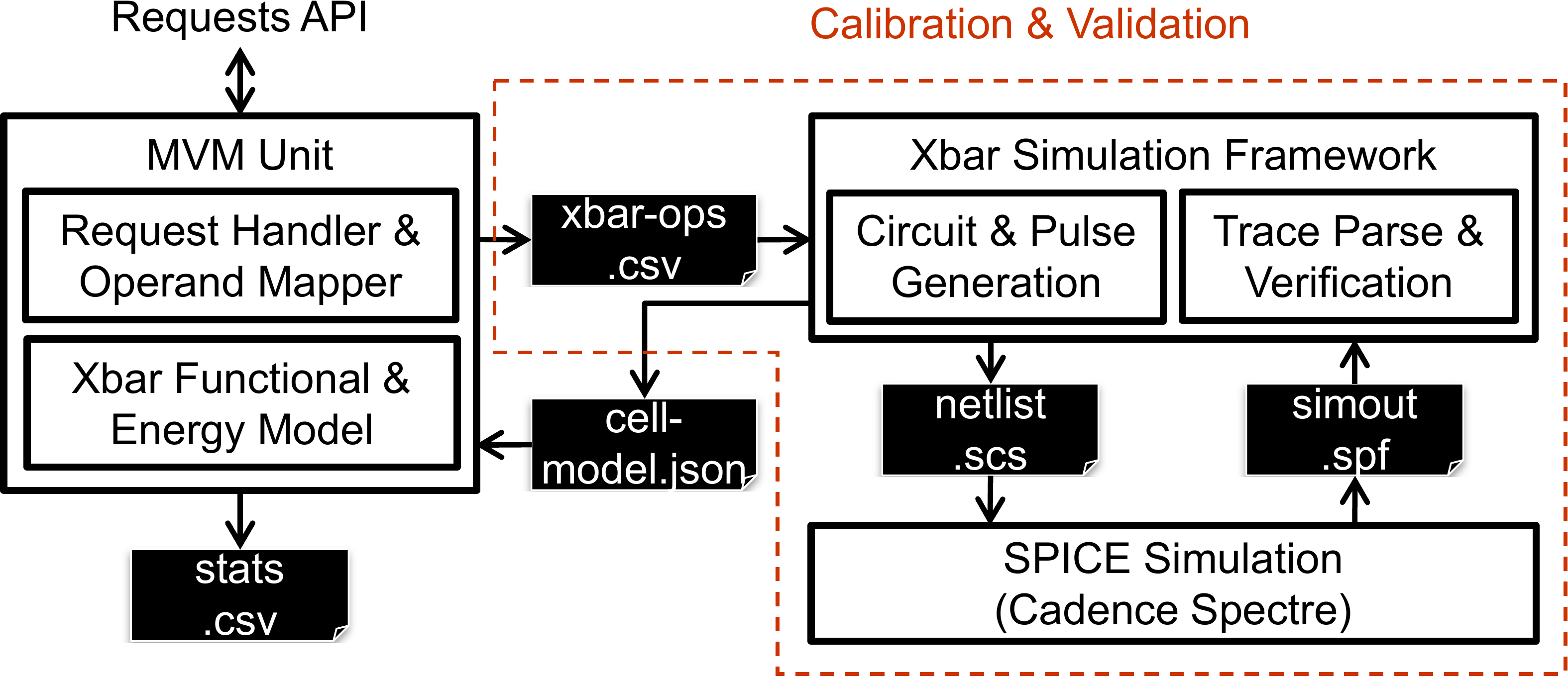}
    \caption{Tools and Simulation Workflow}
    \label{fig:tools}
\end{centering}
\end{figure}

Fig.~\ref{fig:tools} shows the workflow and tools developed in this work.
The \textit{MVM Unit}, shown on the left side, implements the fast energy model.
It receives MVM requests using signed or unsigned integer operands with a resolution of up to 16 bits.
The incoming requests are translated into a set of \textit{xbar-ops}, i.e.\ single pulse operations using the data types the target crossbar requires.
This transformation includes mapping weights to conductances and applying bit-slicing when necessary.
The \textit{xbar-ops} are utilised by the \textit{Xbar Functional \& Energy Model} to generate the MVM output vector.
Simultaneously, the \gls{mvm} energy cost is computed using Eq.~\ref{eq:emvm_short} and the calibrated cell parameters in \textit{cell-model.json}.

When adding a new cell configuration, a calibration process is necessary.
It consists of transient SPICE simulations of the single cell, pulse-based \gls{mac} operation described in Section~\ref{sec:cell_pulse_energy}.
A sweep over the conductance range of the target memristive device is done to extract the parameters $G_{C,min}$, $G_{C,max}$, $\alpha$, and $P_{WL}$.
Finally, the new cell's calibration parameters are stored in \textit{cell-model.json}.

The SPICE simulations are done with Cadence Spectre.
The memristive device is simulated using the \textit{JART VCM-1B} \cite{cueppers_jart_vcmv1b}
compact model, which has been fitted to a \gls{vcm}-type $Pt/HfO_2/TiO_x/Ti$ cell.
For the transistors, we use the PTM models from~\cite{PTM_2006}.

\section{Results and Discussion}

We select four 1T1R cell configurations to demonstrate our calibration methodology.
These are compiled in Table~\ref{tab:1t1r_configs}.
The memristor model parameters from~\cite{cueppers_jart_vcmv1b} are left unchanged, and
only the transistor geometry and wire parasitics are modified.
The wire parasitics' values are based on the estimations for \SI{32}{\nano\meter} technology in the \gls{itrs0}~\cite{ITRS2013}.
The same pulse settings are used for the calibration of all cells.
$V_{RB}$ is set to \SI{0.2}{\volt} and $V_{RW}$ to \SI{1.2}{\volt}.
The common pulse length $T$ is \SI{10}{\nano\second}, with \SI{4}{\nano\second} active time ($T_A$) and \SI{1}{\nano\second} rise-fall time ($T_{RF}$).

1000 \glspl{mvm} are simulated on a 64$\times$64 crossbar for each cell configuration to validate the energy model.
The crossbar is initialised with random 8-bit unsigned integers with a rectified normal distribution.
Binary input vectors with varying sparsity are provided as inputs.
Compared to SPICE, our simulations are up to 1000$\times$ faster and yield energy estimations with an absolute error below 1\% in all cases.

\begin{table}
\begin{centering}
\caption{\label{tab:1t1r_configs}1T1R Array Configurations}
\begin{tabular}{|l|l|ll|ll|ll|}
    \hline
    \multirow{2}{*}{ID} & Transistor         & \multicolumn{2}{l|}{Wire Parasitics} & \multicolumn{2}{l|}{Conductance/\mSIUuS}        & \multicolumn{2}{l|}{Power/\mSIUfJ}  \\ \cline{3-8}
                        & W$\times$L/\mSIUnm & \multicolumn{1}{l|}{C/\mSIUfF}       & R/$\Omega$ & \multicolumn{1}{l|}{Min}  & Max    & \multicolumn{1}{l|}{Min} & Max \\ \hline
    A                   & 32$\times$32       & \multicolumn{1}{l|}{0}               & 0          & \multicolumn{1}{l|}{8.89} & 107.77 & \multicolumn{1}{l|}{1.69} & 19.64 \\ \hline
    B                   & 100$\times$32      & \multicolumn{1}{l|}{0}               & 0          & \multicolumn{1}{l|}{9.37} & 265.41 & \multicolumn{1}{l|}{1.94} & 48.75 \\ \hline
    C                   & 100$\times$32      & \multicolumn{1}{l|}{2}               & 0          & \multicolumn{1}{l|}{9.37} & 265.41 & \multicolumn{1}{l|}{5.32} & 52.13 \\ \hline
    D                   & 100$\times$32      & \multicolumn{1}{l|}{2}               & 2.215      & \multicolumn{1}{l|}{5.60} & 178.83 & \multicolumn{1}{l|}{7.54} & 20.17 \\ \hline
    \end{tabular}
\end{centering}
\end{table}

\subsection{Cell Energy Budget} \label{sec:res_cell_energy}

\begin{figure}
\begin{centering}
        \includegraphics[width=\linewidth]{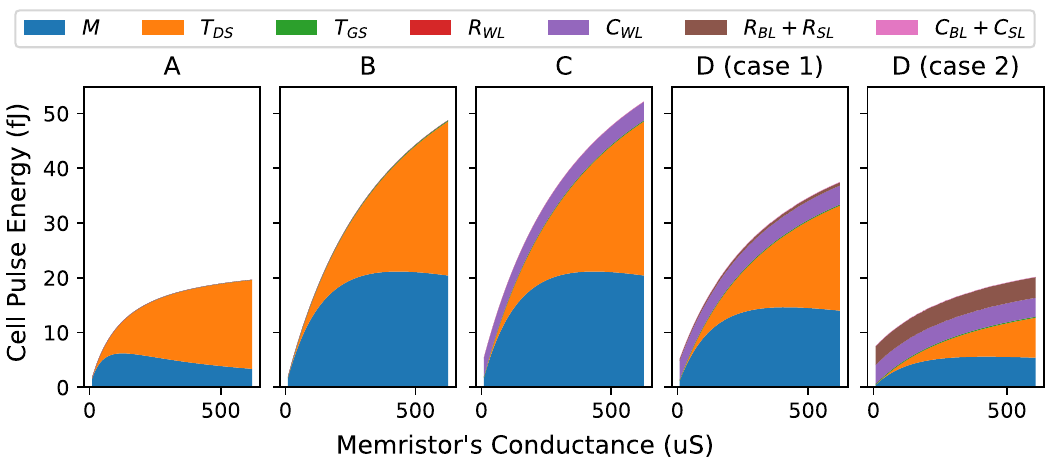}
    \caption{Cell Energy Distribution}
    \label{fig:cell_energy}
\end{centering}
\end{figure}

The energy contribution of all elements in the 1T1R cell against the total cell conductance $G_{C}$ is plotted in Fig.~\ref{fig:cell_energy}.
Configuration A uses a transistor with a geometry ratio $W/L$ of 1, with a high on-state resistance $R_{TON}$ and thus a low cell conductance range.
Configuration B reduces the transistor's voltage drop by using a larger $W/L$.
This change is done to broaden the read-out range, as suggested by~\cite{Yu_ReviewCimChipsDL2021}.
In our case, this leads to an increase of 2.48$\times$ in the total energy.
Configurations C and D illustrate the effect of wire parasitics.
The calibration of configuration D requires a complete crossbar simulation.
We consider a 64$\times$64 crossbar and two cases.
In case 1, all other cells are programmed to $G_{max}/16$, while in case 2, other cells are set to $G_{max}/2$.
A drastic change in the cell energy profile is observed in case 2, as the parasitic resistance dominates the power.

For simplicity, only configuration C is used for the experiments in the following sections.

\subsection{Comparing Data Representations} \label{sec:res_data_rep}

We use our fast model to compare the energy cost of an 8-bit MAC operation for several cell bit-widths and weight mappings.
This experiment uses synthetically generated data for the MVM weights and input vectors.
The input vectors are uniformly distributed 8-bit unsigned integers.
The signed weights are generated using a normal distribution centred at zero. The standard deviation is increased in powers of two.

\begin{figure}
\begin{centering}
    \includegraphics[width=\linewidth]{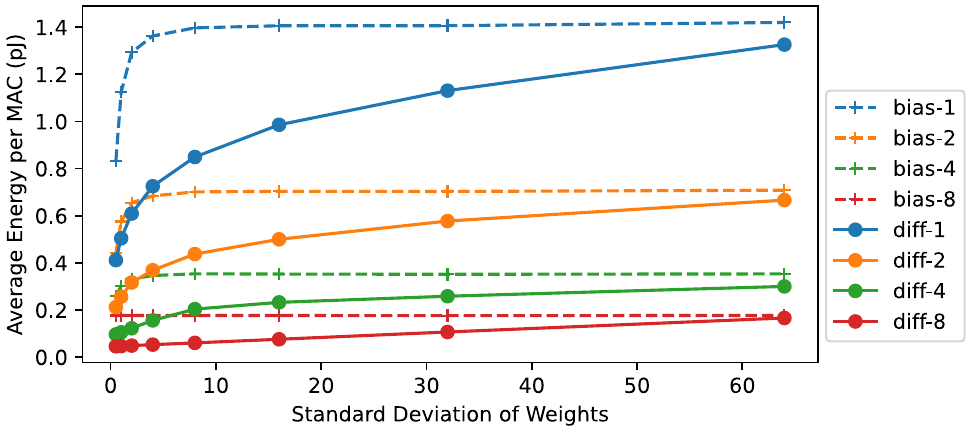}
    \caption{Average Energy per 8-bit MAC Operation with Synthetic Data}
    \label{fig:mappings_synthetic}
\end{centering}
\end{figure}

The average energy per 8-bit MAC operation is depicted in Fig.~\ref{fig:mappings_synthetic}.
Differential mapping proves superior to the bias method, with a more noticeable distinction when the weights' standard deviation approaches zero.

\subsection{Crossbar MVM Energy for CNN Workloads} \label{sec:res_dnn}

\begin{figure}
\begin{centering}
        \includegraphics[width=\linewidth]{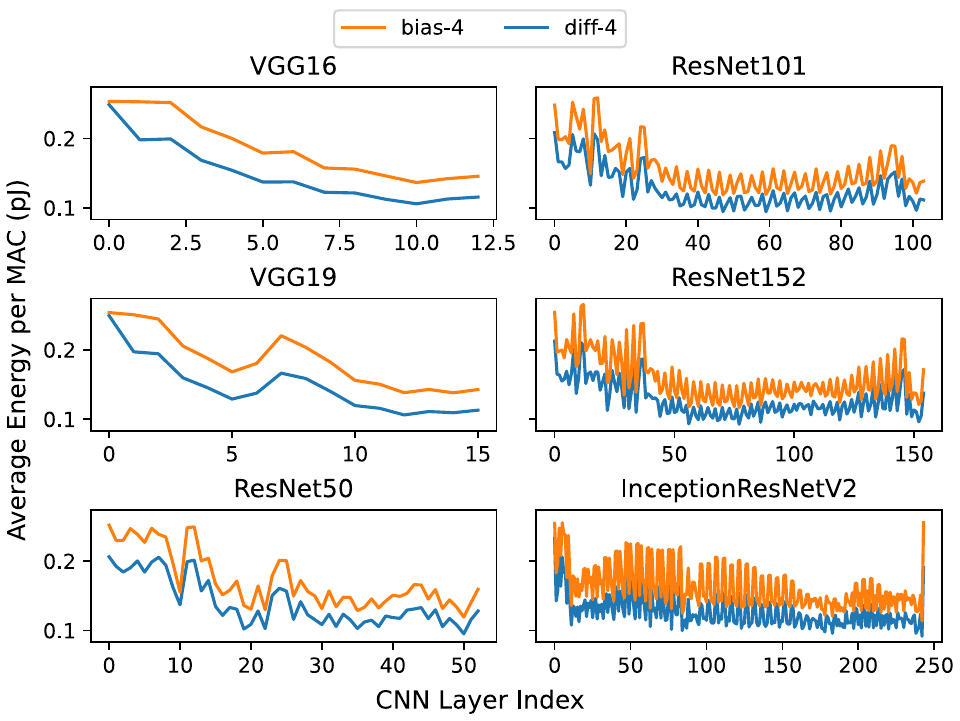}
    \caption{Average Energy per 8-bit MAC Operation for CNN Workloads}
    \label{fig:dnn_energy}
\end{centering}
\end{figure}

The experiment with synthetic data helps compare data representations for the weights mapped in the crossbar cells.
However, it does not show the influence of the input and output data of the MVM operations.
To achieve this, we run experiments with actual weights and input data.

First, we obtained a set of pre-trained \glspl{cnn} from the \textit{Keras} applications database.
We apply \textit{tensorflow-lite} 8-bit quantisation~\cite{tflite_int_only_quant} configured for signed weights and unsigned activations.
We run 100 inference cycles with each \gls{cnn} model using 100 images from the \textit{imagenet} dataset.
For all 2-dimensional convolutions (\textit{Conv2D}), we apply the \textit{im2col} transformation and 
generate energy statistics using our tool.

We compute the average energy cost per 8-bit \gls{mac} operation for each convolutional layer.
The results are plotted in Fig.~\ref{fig:dnn_energy}.
For simplicity, we only show the results obtained with 4-bit cells.
The smaller networks, VGG16 and VGG19, show a decline in energy as the data travels towards the final layers.
This pattern is also noticeable in the more complex networks Resnet50, Resnet101, and InceptionResNetV2.
However, in the latter case, the required energy oscillates.
Consistent with the observation in the previous section, the differential mapping outperforms the bias mapping in all cases.

\section{Related Work}

The authors of \cite{CIM_SRAM_MICAS_2023} propose energy models for analogue SRAM-based CIM kernels.
A detailed study on the performance and energy of 1T1R RRAM crossbars as storage media is presented in ~\cite{1T1R_perf_models}.
The work in \cite{sandia_perf_estimation_JETCAS2018} extensively analyses RRAM-based CIM kernels but only considers diodes as selection devices.

MNSIM~\cite{MNSIM_2016} and NeuroSim~\cite{NeuroSim_2017} are two frameworks that offer energy statistics for RRAM CIM kernels,
utilizing intricate circuit models with numerous configurable parameters.
In contrast to our approach, none of these tools considers the impact of wire parasitic resistance on crossbar energy.
Furthermore, our calibration-based method condenses circuit-level details into just a few parameters,
enhancing simulation speed and facilitating system-level analysis.

\section{Conclusion}

RRAM crossbars are at the core of numerous analogue CIM concepts for \gls{dnn} acceleration.
Assessing their energy efficiency requires a cross-layer approach considering circuit implementation details and workload properties.
This work presents a methodology to achieve this through calibrated abstract models.
Our analysis of state-of-the-art weight mapping schemes shows a clear advantage of differential-based schemes.
Future work will extend our modelling approach to evaluate alternative circuit topologies, quantisation, and data representation schemes.

\input{main_arxiv.bbl}

\end{document}